# Improved Analytical/Statistical Modelling of the Shock Wave-Laminar Boundary Layer Around a Thin Airfoil: Standard Atmosphere Case

Nasser Eddegdag[1], Omar El-Aajine[2], Aze-eddine Naamane[3], Mohammed Radouani[4]

[1] Moulay Ismail University, Meknes, Morocco

n.eddegdag@edu.umi.ac.ma

[2] Moulay Ismail University, Meknes, Morocco

o.elaajine@edu.umi.ac.ma

[3] Royal Moroccan Air School, Marrakech, Morocco

azeddine.naamane@gmail.com

[4] Moulay Ismail University, Meknes, Morocco

m.radouani@ensam-umi.ac.ma

**Abstract.** The aim of this present work is to develop an improved and more precised analytical modelling of a steady irrotational laminar Shock Wave-Boundary Layer Interaction for weak shockwaves around a thin airfoil at a low incidence in the standard atmosphere. This study adapts the asymptotical modelling of our problem treated by our research team previously [1], and improves the analytical resolution process by integrating the empirical parameter $m$. Then, confrontation of our analytical model to experimental results obtained through experimentation in Supersonic Wind Tunnel AF300. And finally, a CFD numerical simulation in ANSYS Fluent R13 was conducted in order to validate our model, followed up by a statistical study in SPSS taking into consideration analytical and numerical results in order to establish the exact analytical expression of $m$ for each airfoil depending on upstream Mach numbers .The comparison of results obtained from the numerical, experimental and analytical confrontations of our improved and previous models showed promising results and greatly decreased the approximation errors obtained in our previous studies.

# 1   Introduction

Because of their ubiquitous presence in high-speed flights and their impact on vehicles and components performances Shock Wave- Boundary Layer Interaction "SWBLI" has been studied for more than 50 years [2]. Despite remarkable progress in computational and measurements capabilities, there are more important quantities that cannot be predicted very accurately. In high Reynolds number flows, the no-slip condition at solid surfaces leads to the formation of a boundary layer in which the velocity and the temperature adjust rapidly to meet the local freestream conditions. Hence, a shockwave is a localized pressure increase and is classified as an adverse pressure gradient. The effects of a moderate adverse pressure gradient are a thickening of a boundary layer [3]. But in extreme cases, where the shockwave is strong enough and large enough, it causes boundary layer separation, which then has a global effect on the structure of the flow and hence the solid. The study, knowledge and understanding harvested from SWBLI studies is vital given the omnipresent nature of SWBLI in high-speed flights. Consider the example of two missiles flying in close proximity to each other or that of the side intake with central bodies of a supersonic aircraft, Wing-Body interactions and static props in the presence of a shockwave are all applications in which SWBLI plays the main role [4].

In our previous studies, we were interested in the modelling of laminar steady irrotational SWBLI around a thin airfoil located in the standard atmosphere. The modelling process initiated from Navier-Stokes equations adapted to our assumptions, to which we applied the asymptotic methods, adimensionnalisation and linearization, in order to obtain the equation governing our problem [1].

Afterwards, we applied an asymptotical expansion on the governing equation following $\beta$ and $Re^{-1}$ individually at $o(1)$. With $\bar{U}_\infty$, characterizing the dimensionless upstream velocity, which in our case in the standard atmosphere, for which the solution is considered simple since we consider that the upstream flow is slightly disturbed by the thin airfoil, irrotational and varies linearly following $\bar{\eta}$ which is considered the only variable characterizing the upstream velocity. In addition, we applied the singular perturbations methods in respect to the principle of least degeneracy, in order to find expression forms of perturbation velocity potential $\varphi$ near the airfoil body far from the leading edge, both inside the boundary layer $\hat{\varphi}_{01}(\bar{s},\hat{\eta})$ and in the undisturbed flow outside the boundary layer $\bar{\varphi}_{00}(\bar{s},\bar{\eta})$ [5][6].

The resolution process of our past work consisted of adopting the same technics (Triple Deck Technic) used in this present work [1], however the linkage methods between the different decks of the boundary layer, especially the linkage between the sub-viscous deck and the main deck where the logic adopted was to connect the two layers based on the continuity of the relative thickness that tends in each deck to meet an intermediate variable, neglecting the effect of connectivity in velocity components are what caused slight lack of precision in our previous model [7]. Mach numbers in each layer tend to be continuous and converge in each layer to an intermediate variable *m* which is the local incompressible connection Mach number that characterizes the continuity of velocity (continuity of variables and their derivatives) and the unique incompressible aspect of the sub-viscous deck. The neglection of this empirical parameter has effects on the formulation of the velocity field expressions inside and outside the boundary layer since the resolution methods to find exact expressions of constants link all the expressions together, hence inserting and taking into consideration the effect of that parameter will minimize errors in Mach number expressions in the different study parts of our problem.

## 2      Resolution Process

The resolution process of our problem focalizes on finding exact expressions of constants in the formulas of $\bar{\varphi}_{00}$ and $\bar{\varphi}_{01}$, in order to establish exact formulas of Mach numbers inside and outside the boundary layer near the wall far from the leading edge of the thin airfoil. The technic adopted in our process is the "triple-layer" model, which was proposed in 1969, independently and almost simultaneously, by NEILAND (1969) [8] and STEWARTSON and WILLIAMS (1969) [9], makes it possible to account for the behavior of the boundary layer (here it will be of a laminar boundary layer) in various circumstances where it undergoes "an accident" in the vicinity of a line drawn on an obstacle. Naturally, we are interested in the local structure of the flow when $Re \to \infty$ and we are then led to locally distinguish three layers (triple deck) (figure 1): ***Upper deck***, ***Main deck*** and ***Lower deck***

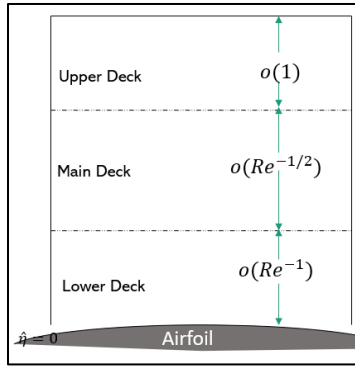

**Fig. 1. The triple deck model of NEILAND and STEHARTSON-WILLIAMS**

### 2.1      The Connection Between the Main Deck and The Lower Deck (Link Between the Incompressible and The Compressible)

Near the wall away from the leading edge when $\bar{\eta} \to 0$, the steady flow is considered that of a low viscosity and homogeneous incompressible fluid. Thus, we have the equation of the velocity field for a low viscous incompressible flow, from which we have the equation of the velocity field in the viscous sub-layer [5] of thickness $\tilde{\eta}$ relative to:

$$\begin{cases} \widetilde{U}^{LD}(\bar{s},\hat{\eta}) = \dfrac{\bar{U}_\infty(0)\cos(\alpha)}{3}\tilde{\eta}^2 \\ \widetilde{V}^{LD}(\bar{s},\tilde{\eta}) = 0 \end{cases} \quad (1)$$

Where $\alpha$ is the angle of incidence, $\bar{\eta}$ and $\bar{s}$ are normal and tangential coordinate in the Frenet coordinate system $(P,\vec{e},\vec{n},\vec{b})$, $\widetilde{U}^{LD}$ and $\widetilde{V}^{LD}$ are the dimensionless longitudinal and transverse velocity in the Lower deck, with $\hat{\eta}$ is the normal coordinate (dimensionless) in the vicinity of the wall in the Main deck and $\tilde{\eta}$ is the normal coordinate (dimensionless) in the vicinity of the wall in the Lower deck.

Thus, we are interested in the area of connection between the Lower deck and the Main deck (figure 2) when $\tilde{\eta} \to \tilde{\eta}^*$ and $\hat{\eta} \to \hat{\eta}^*$ with $\tilde{\eta} = \dfrac{\hat{\eta}}{Re^{-p}}$, with $(Re^{-1})^p$ a gauge function a priori unknown and they must be determined so that the model in triple layer, set up is asymptotically, when $Re^{-1} \to 0$, consistent [10]. As well as the Local number of Mach tends towards an incompressible Mach, hence $M_{local} \to m$ with $0 \prec m \prec 0.3$.

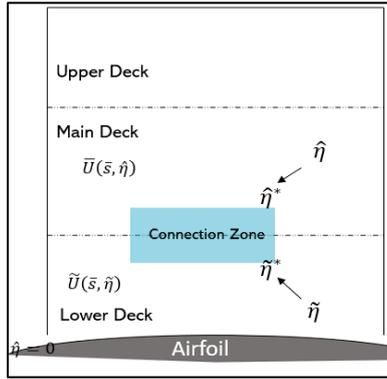

**Fig. 2. Diagram showing the connection between the Lower deck and the Main deck**

Thus, by taking the normal component of the connection between the Main deck and Lower, we have:

$$\hat{V}_{MD}(\bar{s},\hat{\eta}) = \widetilde{V}^{LD}(\bar{s},\tilde{\eta}) \Rightarrow A = B\dfrac{2\tan\left(\dfrac{m^2}{2}\pi\right)}{\sqrt{\dfrac{m^4}{1-m^2}}} \quad (2)$$

And by taking the tangential component of the connection we have: $\widehat{U}_{MD}(\bar{s},\hat{\eta}) = \widetilde{U}^{LD}(\bar{s},\tilde{\eta})$, and so we get

$$A = \dfrac{1.7783\, m\cos(\alpha)}{\cos\left(\dfrac{m^2}{2}\pi\right)\left(\dfrac{m^5}{4}+\dfrac{m^3}{2}\right)\left(\dfrac{m^4}{1-m^2}\right)} Re^{2p+1} \quad (3)$$

With $p = -1$, since the constants of the perturbed potentials must be of order $o(1)$, hence we will have a condition on $p$ which will manage the link between $\hat{\eta}^*$ and $\tilde{\eta}^*$. And so, the connection zone is of thickness $l_0 Re^{-1}$ (where $l_0$ is the length of the profile wall) [11]. But, since $A$ is of the order $o(1)$, thus, we find that for $1 \leq A \leq 5$ we must have $0.153 \leq m \leq 0.2$. Hence, we have the exact expression of non-disturbed viscous non-dimensional velocity potential in the Main deck:

$$\hat{\varphi}_{01}(\bar{s},\hat{\eta}) = \begin{bmatrix} 0.77\left(\frac{M_\infty^5}{4} + \frac{M_\infty^3}{2}\right) \\ \sqrt{\frac{M_\infty^4}{M_\infty^2-1}} \frac{m\cos(\alpha)}{\cos\left(\frac{m^2}{2}\pi\right)\left(\frac{m^5}{4}+\frac{m^3}{2}\right)\left(\frac{m^4}{1-m^2}\right)} Re^{-2} \\ \left(\cos\left(\frac{M_\infty^2}{2}\pi\right)\sin\left(\frac{1}{2}\sqrt{\frac{M_\infty^4}{M_\infty^2-1}}\bar{s}\right) - \right. \\ \left. \frac{\sin\left(\frac{M_\infty^2}{2}\pi\right)}{2\tan\left(\frac{m^2}{2}\pi\right)}\sqrt{\frac{m^4}{1-m^2}}\cos\left(\frac{1}{2}\sqrt{\frac{M_\infty^4}{M_\infty^2-1}}\bar{s}\right)\right) \end{bmatrix} \hat{\eta}^2 + D_3 \quad (4)$$

With $D_3$ being a constant.

## 2.2 The Connection Between the Main Deck and The Upper Deck (Link Between Boundary Layer and Exterior Supersonic Flow)

We have the connection between the Upper layer and the Main layer is characterized by the sonic line which characterizes the passage from the subsonic flow to a supersonic flow in $\bar{\varphi}_{00}(\bar{s},\bar{\eta})$ (figure 3), thus the connection zone is characterized by a transonic flow in which $M_\infty^2 = 1 \Rightarrow \bar{U}^2 + \bar{V}^2 = 1$

Thus, in this zone where;

$$\begin{pmatrix} \hat{\eta} \to \hat{\eta}^* \text{ et } \bar{\eta} \to \bar{\eta}^* \text{ with } \hat{\eta}^* = \frac{\bar{\eta}^*}{Re^{-q}} \\ \text{also } \bar{s} \to \bar{s}^* \end{pmatrix} \quad (5)$$

We suppose that the curve of the sonic line is characterized by a vertex $S(\bar{s}^*_0, \hat{\eta}^*_0)$ [10][12] in which:

$$\begin{cases} \bar{V}_{MD}(\bar{s}^*_0, \hat{\eta}^*_0) = 0 \\ \bar{V}_{ext}(\bar{s}^*_0, \bar{\eta}^*_0) = 0 \end{cases} \& \begin{cases} \bar{U}_{MD}(\bar{s}^*_0, \hat{\eta}^*_0) = \delta_0 \\ \bar{U}_{ext}(\bar{s}^*_0, \bar{\eta}^*_0) = \delta_0 \end{cases} \text{ with } \delta_0 = \pm 1 \quad (6)$$

So, to find the coordinates of the vertex, we solve the system:

$$\begin{cases} \bar{V}_{MD}(\bar{s}^*_0, \hat{\eta}^*_0) = 0 \\ \bar{U}_{MD}(\bar{s}^*_0, \hat{\eta}^*_0) = \delta_0 \end{cases} \quad (7)$$

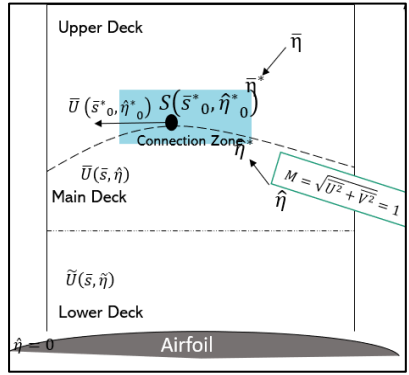

**Fig. 3. Diagram showing the connection between the Main deck and the Upper deck**

Hence, we get

$$\bar{s}^*{}_0 = \frac{2\,arctan\left(\frac{tan\left(\frac{M_\infty^2}{2}\pi\right)}{2\,tan\left(\frac{m^2}{2}\pi\right)}\sqrt{\frac{m^4}{1-m^2}}\right)}{\sqrt{\frac{M_\infty^4}{M_\infty^2-1}}} \quad (8)$$

Although, since $0 \leq \bar{s}^*{}_0 \leq 1$, by plotting $\bar{s}^*{}_0$ as a function of $M_\infty$, we notice that our model is valid for a supersonic Mach as $1.41 \leq M_\infty \leq 1.6864$. Also, we have

$$\bar{U}_{MD}(\bar{s}^*{}_0, \hat{\eta}^*{}_0) = \frac{1}{1-\hat{\eta}^*{}_0}\frac{\partial \bar{\varphi}_{01}(\bar{s},\hat{\eta})}{\partial \bar{s}} = \delta_0 = \pm 1 \quad (9)$$

Hence

$$\hat{\eta}^*{}_0 = \frac{1-\sqrt{1-4\tau}}{2\tau} \text{ with } \tau = \left[\frac{\frac{0.385 m\,cos(\alpha)}{cos\left(\frac{m^2}{2}\pi\right)\left(\frac{m^5}{4}+\frac{m^3}{2}\right)\left(\frac{m^4}{1-m^2}\right)}Re^{-1}\left(\frac{M_\infty^5}{4}+\frac{M_\infty^3}{2}\right)\left(\frac{M_\infty^4}{M_\infty^2-1}\right)}{cos\left(\frac{M_\infty^2}{2}\pi\right)cos\left(\frac{1}{2}\sqrt{\frac{M_\infty^4}{M_\infty^2-1}}\bar{s}^*{}_0\right)} \\ +\frac{sin\left(\frac{M_\infty^2}{2}\pi\right)}{2\,tan\left(\frac{m^2}{2}\pi\right)}\sqrt{\frac{m^4}{1-m^2}}sin\left(\frac{1}{2}\sqrt{\frac{M_\infty^4}{M_\infty^2-1}}\bar{s}^*{}_0\right)\right] \quad (10)$$

And so, after finding the coordinates for the vertex $S(\bar{s}_0, \hat{\eta}_0)$, we resolve the following system (11) in order to find the exact expression of the undistributed inviscid dimensionless velocity potential in the Upper deck and hence find the values of the constants in $\bar{\varphi}_{00}(\bar{s}, \bar{\eta})$, we have:

$$\bar{\varphi}_{00}(\bar{s},\bar{\eta}) = (1-\bar{\eta})^{\frac{-M_\infty^2}{2}}\left[cos\left(\frac{M_\infty^2}{2}\pi\right)(C_1+C_2)cos\left(\frac{1}{2}\sqrt{\frac{M_\infty^4}{M_\infty^2-1}}\bar{s}\right) \\ + sin\left(\frac{M_\infty^2}{2}\pi\right)(C_1-C_2)sin\left(\frac{1}{2}\sqrt{\frac{M_\infty^4}{M_\infty^2-1}}\bar{s}\right)\right]D \quad (11)$$

We put $\begin{cases} Z = C_1 + C_2 \\ X = C_1 - C_2 \end{cases}$

In the connection zone between the Upper and Main deck we have: $\bar{\eta}^*{}_0 = Re^{-q}\hat{\eta}^*{}_0$, although since $0 \leq \bar{\eta}^*{}_0 \leq 1$, then we conclude that $\bar{\eta}^*{}_0 = \hat{\eta}^*{}_0$.

To find the values of constants $Z$ and $X$, we resolve the following equations:

$$\bar{V}_{ext}(\bar{s}^*{}_0, \bar{\eta}^*_0) = 0 \Rightarrow Z = -X\frac{tan^2\left(\frac{M_\infty^2}{2}\pi\right)}{2\,tan\left(\frac{m^2}{2}\pi\right)}\sqrt{\frac{m^4}{1-m^2}} \quad (12)$$

$$\bar{U}_{ext}(\bar{s}^*{}_0, \bar{\eta}^*_0) = \bar{U}_{MD}(\bar{s}^*{}_0, \bar{\eta}^*_0) = -1 \Rightarrow DX = -\frac{2}{(1-\bar{\eta}^*_0)^{\frac{-M_\infty^2}{2}}\sqrt{\frac{M_\infty^4}{M_\infty^2-1}}sin\left(\frac{M_\infty^2}{2}\pi\right)\chi} \quad (13)$$

with

$$\chi = \left[\frac{tan\left(\frac{M_\infty^2}{2}\pi\right)}{2\,tan\left(\frac{m^2}{2}\pi\right)}\sqrt{\frac{m^4}{1-m^2}}sin\left(arctan\left(\frac{tan\left(\frac{M_\infty^2}{2}\pi\right)}{2\,tan\left(\frac{m^2}{2}\pi\right)}\sqrt{\frac{m^4}{1-m^2}}\right)\right) \\ + cos\left(arctan\left(\frac{tan\left(\frac{M_\infty^2}{2}\pi\right)}{2\,tan\left(\frac{m^2}{2}\pi\right)}\sqrt{\frac{m^4}{1-m^2}}\right)\right)\right] \quad (14)$$

Accordingly, we have:

$$\bar{\varphi}_{00}(\bar{s},\bar{\eta}) = \frac{2(1-\bar{\eta})^{\frac{-M_\infty^2}{2}}}{(1-\bar{\eta}_0^*)^{\frac{-M_\infty^2}{2}}\sqrt{\frac{M_\infty^4}{M_\infty^2-1}}\chi} \left[ \frac{tan\left(\frac{M_\infty^2}{2}\pi\right)}{2\,tan\left(\frac{m^2}{2}\pi\right)} \sqrt{\frac{m^4}{1-m^2}} cos\left(\frac{1}{2}\sqrt{\frac{M_\infty^4}{M_\infty^2-1}}\bar{s}\right) - sin\left(\frac{1}{2}\sqrt{\frac{M_\infty^4}{M_\infty^2-1}}\bar{s}\right) \right] \quad (15)$$

## 3     Results For Disturbance Potential and Mach Number

The application of the asymptotic method in the equations, both in the Navier-Stocks model and in the boundary conditions, gave two systems predicted by the asymptotic expansion of $\bar{\varphi}_0$. according to the generalized development of $\bar{\varphi}_0$ with respect to $Re^{-1}$[1], we have:

$$\bar{\varphi}_0(\bar{s},\bar{\eta}) = \bar{\varphi}_{00}(\bar{s},\bar{\eta}) + Re^{-1}\bar{\varphi}_{01}(\bar{s},\bar{\eta}) \quad (16)$$

However, thanks to the singular perturbations' methods, the triple deck technique and the boundary conditions, the expressions $\bar{\varphi}_{00}(\bar{s},\bar{\eta})$ have been obtained, so we will have the tangential and normal velocity components outside the boundary layer. From result of the velocity perturbation potential which was obtained in our calculations, we can deduce the expression of the velocity field on the wing profile. According to the linearization theorem or Hartman–Grobman theorem [1], we have:

$$\bar{U}_{ext}(\bar{s},\bar{\eta}) = \bar{U}_\infty(\bar{s},\bar{\eta}) + \frac{\beta}{1-\bar{\eta}}\frac{\partial\bar{\varphi}_{00}(\bar{s},\bar{\eta})}{\partial\bar{s}} \; \& \; \bar{V}_{ext}(\bar{s},\bar{\eta}) = \beta\frac{\partial\bar{\varphi}_{00}(\bar{s},\bar{\eta})}{\partial\bar{\eta}} \quad (17)$$

With $\beta$ the relative thickness of the airfoil, $\bar{U}_\infty(\bar{s},\bar{\eta})$ is the speed of the dimensionless standard atmosphere. Hence, we get:

$$\bar{U}_{ext}(\bar{s},\bar{\eta}) = -1.155 M_\infty(\bar{\eta}-1) + \frac{\beta(1-\bar{\eta})^{\frac{-M_\infty^2}{2}-1}}{(1-\bar{\eta}_0^*)^{\frac{-M_\infty^2}{2}}\chi}K_1 \; \& \; K_1 = \frac{tan\left(\frac{M_\infty^2}{2}\pi\right)}{2\,tan\left(\frac{m^2}{2}\pi\right)}\sqrt{\frac{m^4}{1-m^2}} sin\left(\frac{1}{2}\sqrt{\frac{M_\infty^4}{M_\infty^2-1}}\bar{s}\right) - cos\left(\frac{1}{2}\sqrt{\frac{M_\infty^4}{M_\infty^2-1}}\bar{s}\right) \quad (18)$$

And

$$\bar{V}_{ext}(\bar{s},\bar{\eta}) = \beta\frac{M_\infty^2(1-\bar{\eta})^{\frac{-M_\infty^2}{2}-1}}{(1-\bar{\eta}_0^*)^{\frac{-M_\infty^2}{2}}\sqrt{\frac{M_\infty^4}{M_\infty^2-1}}\chi}K_2 \; \& \; K_2 = \frac{tan\left(\frac{M_\infty^2}{2}\pi\right)}{2\,tan\left(\frac{m^2}{2}\pi\right)}\sqrt{\frac{m^4}{1-m^2}} cos\left(\frac{1}{2}\sqrt{\frac{M_\infty^4}{M_\infty^2-1}}\bar{s}\right) - sin\left(\frac{1}{2}\sqrt{\frac{M_\infty^4}{M_\infty^2-1}}\bar{s}\right) \quad (19)$$

We obtained thanks to the methods of the singular perturbations and the boundary conditions the expression $\hat{\varphi}_{01}(\bar{s},\hat{\eta})$, so we will have the tangential and normal velocity components inside the boundary layer. From the result of the speed disturbance potential that was obtained in our calculations, we can deduce the expression of the speed field on the wing profile. According to the linearization theorem or Hartman–Grobman theorem, we have:

$$\bar{U}_{BL}(\bar{s},\hat{\eta}) = \beta\frac{\partial\hat{\varphi}_{01}(\bar{s},\hat{\eta})}{\partial\bar{s}} \; \text{and} \; \bar{V}_{BL}(\bar{s},\hat{\eta}) = \beta\frac{\partial\hat{\varphi}_{01}(\bar{s},\hat{\eta})}{\partial\hat{\eta}} \quad (20)$$

Hence,

$$\bar{U}_{BL}(\bar{s},\hat{\eta}) = \beta \begin{bmatrix} 0.385 \left(\frac{M_\infty^5}{4} + \frac{M_\infty^3}{2}\right)\left(\frac{M_\infty^4}{M_\infty^2-1}\right) \\ \frac{m\cos(\alpha)}{\cos\left(\frac{m^2}{2}\pi\right)\left(\frac{m^5}{4}+\frac{m^3}{2}\right)\left(\frac{m^4}{1-m^2}\right)} Re^{-2} \\ \left[\cos\left(\frac{M_\infty^2}{2}\pi\right)\cos\left(\frac{1}{2}\sqrt{\frac{M_\infty^4}{M_\infty^2-1}}\bar{s}\right) \\ + \frac{\sin\left(\frac{M_\infty^2}{2}\pi\right)}{2\tan\left(\frac{m^2}{2}\pi\right)}\sqrt{\frac{m^4}{1-m^2}}\sin\left(\frac{1}{2}\sqrt{\frac{M_\infty^4}{M_\infty^2-1}}\bar{s}\right)\right] \end{bmatrix} \hat{\eta}^2 \quad (21)$$

And

$$\bar{V}_{BL}(\bar{s},\hat{\eta}) = \beta \begin{bmatrix} 1.54 \left(\frac{M_\infty^5}{4} + \frac{M_\infty^3}{2}\right)\sqrt{\frac{M_\infty^4}{M_\infty^2-1}} \\ \frac{m\cos(\alpha)}{\cos\left(\frac{m^2}{2}\pi\right)\left(\frac{m^5}{4}+\frac{m^3}{2}\right)\left(\frac{m^4}{1-m^2}\right)} Re^{-2} \\ \left[\cos\left(\frac{M_\infty^2}{2}\pi\right)\sin\left(\frac{1}{2}\sqrt{\frac{M_\infty^4}{M_\infty^2-1}}\bar{s}\right) \\ - \frac{\sin\left(\frac{M_\infty^2}{2}\pi\right)}{2\tan\left(\frac{m^2}{2}\pi\right)}\sqrt{\frac{m^4}{1-m^2}}\cos\left(\frac{1}{2}\sqrt{\frac{M_\infty^4}{M_\infty^2-1}}\bar{s}\right)\right] \end{bmatrix} \hat{\eta} \quad (22)$$

Our main goal is to find the local Mach number on the wall far from the leading edge of the wing airfoil. From this number, we can deduce all the other physical quantities: pressure, density, temperature, flow speed … a formula was useful for us to carry out our calculations:

$$M_{local} = \sqrt{[\bar{U}(\bar{s},\bar{\eta})]^2 + [\bar{V}(\bar{s},\bar{\eta})]^2} \quad (23)$$

With $1.41 \leq M_\infty \leq 1.6864$

Hence, we obtain the core result of our study which is the Local Mach Number around the airfoil, which characterizes all the governing parameters that play a substantial factor in the control and definition of the behavior of the supersonic flow of a laminar viscous fluid around a thin airfoil.

## 4 Validation by Numerical Simulation

In this part, the numerical approach for the study of the problem will be approached, under the same assumptions as in the theoretical approach. Indeed, it will make it possible to obtain the precise behaviour of a viscous laminar supersonic airflow around the NACA 43013 airfoil after setting the simulation software Ansys Fluent R19.3.

### 4.1 Modelling of the problem in Ansys Fluent R19.3

The modelling of the problem on the Ansys fluent software requires a specific study and a great precision in the different stages of the numerical simulation, the study geometry is airfoil NACA 43013, the calculation domain and meshing were done similar to pervious work by our research team [6], taking into account a hyper fine and precise mesh in the vicinity of the edge of the airfoil in order to be able to model the different sub-layers of the boundary layer. Indeed, we have achieving *104777* nodes and *103477* elements stacking in the direction normal to the boundary using a feature called "Inflation" (figure 4)

N. Eddegdag, O. El-Aajine, A. Naamane, M. Radouani                                    9

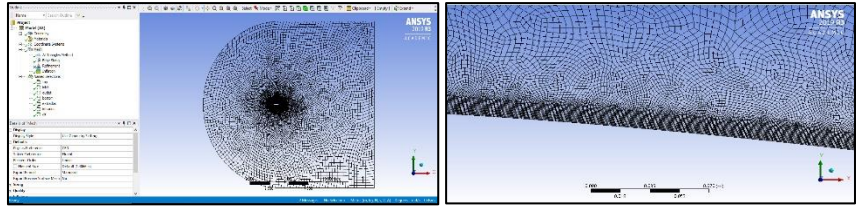

**Fig. 4. Final calculation domain for mesh**

For laminar, steady and compressible flow and viscous fluid, we consider a Reynolds number $Re = 3.42 * 10^5$ and with a free upstream velocity of $M_\infty = 1.45$ and $M_\infty = 1.5$. The fluid (air), at room temperature 288.16 K°, has a density 1.225 kg/m³ and a dynamic viscosity $1.789 * 10^5 kg.(m.s)^{-1}$ [13].

### 4.2   Visualization and analysis of results for NACA 43013

For $M_\infty$=1.45, Figure 5 below is obtained. First of all, it allows to visualize the variation of the Mach number from the leading edge and all along the airfoil. In particular, it is observed that on the upper surface the Mach number keeps increasing from the leading edge to the trailing edge where $M_{local}{}^1$=1.666. While on the lower surface, it grows faster from the leading edge to the trailing edge where $M_{local}{}^2$=1.616. Secondly, the figure clearly depicts the curved detached shock wave on the leading edge, with the presence of a blue-green stagnation sonic bubble where the flow is locally rotational unsteady and incompressible and $M$<1.00, we can also observe at the end of the extrados leading edge where the shape of the airfoil is more of a convex corner the formation of infinite Mach lines that are the result of Prandtl-Meyer expansion waves, same phenomena occurs in the middle of the intrados part of the airfoil near the war, and finally we have the weak oblique SW formed near the intrados part of the leading edge where the airfoil takes the form of a concave corner [14].

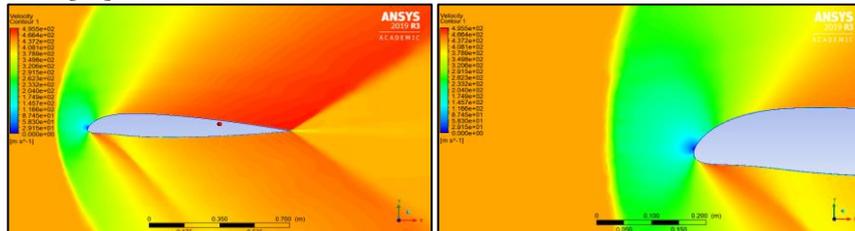

**Fig. 5. Velocity evolution around the NACA 43013 wing airfoil, simulation for $M_\infty$=1.45**

Finally, the third observed phenomenon is the SWBLI, which is the result of viscous supersonic flow. According to figure 6, we note the existence of a boundary layer all along the wall of the thickness profile $l_0$=0.4076mm where the Mach is subsonic and the speed tends towards a zero-value approaching the surface of the profile, at the end of the boundary layer all around the airfoil we have Mach lines [*expansion waves- weak SW*] that are incident shocks towards the boundary layer, taking into consideration for our case their weak nature causing the SWs to go through the boundary layer without changing its global aspect and to slightly locally separate, all over the extrados, its Lower Deck while overall maintaining its same total width. [*the width of the different decks changes imperceptibly in many points of the airfoil, in a pseudo-periodic nature, while maintaining the same total width of the boundary layer*]

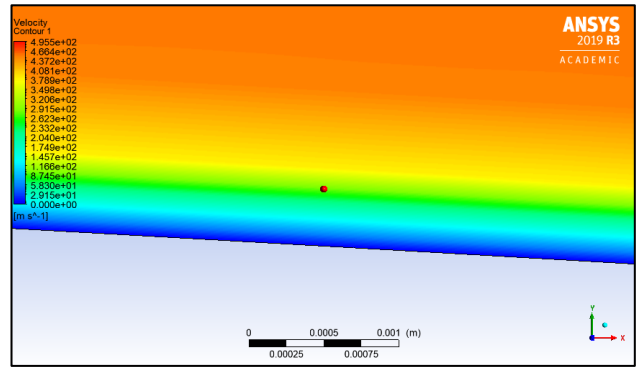

**Fig. 6. Visualization of the boundary layer around the profile by Simulation on Ansys Fluent- in dark blue Lower Deck, in light blue Main Deck, in light green Upper Deck-**

## 5 Confrontation of Numerical-Analytical Results

The various theoretical studies carried out since the advent of the supersonic cover large areas and make it possible to compare several parameters. Our focus on our study is to compare the velocity evolution by comparing numerical and analytical results of our model of local Mach numbers near airfoil wall.

### 5.1 Comparison of Numerical-Analytical results

We took results in the wall far from the leading edge; hence we took results from an $x_0$ which characterizes the end of the leading edge, then for the incompressible Mach m which characterizes the connection between the viscous underlayer and the intermediate layer, we decided to fix its value thanks to the numerical results to have a better precision and a weak error. Local Mach extrados and intrados located in the wall far from the leading edge, thus the position of the maximum relative camber characterizes the end of the leading-edge zone. We perform our calculations and extract our results for x greater than *0.15m*. Similarly, we have according to the theoretical model that $0 \leq \bar{\eta} \leq 0.13257143$, from which we have y limited in 0 and 0.19 for x included between 0.15 and 0.208866, as well as y limited by 0 and 0.132 for x between 0.208866 and 1. It can be seen from figure 7 that for x in the first half of the extrados, up to 40% of the chord, the error is very minimal, almost $\varepsilon^1{}_{ectrados} = 2.283977\%$, while approaching the trailing edge the error increases to a value of $\varepsilon^2{}_{ectrados} = 10.10676\%$, since the region of $x \in [0.2088,1]$ is an instability region with $[0.2088,0.32]$ a transition region which is the response of the oscillation that perturbs the surrounding flow.

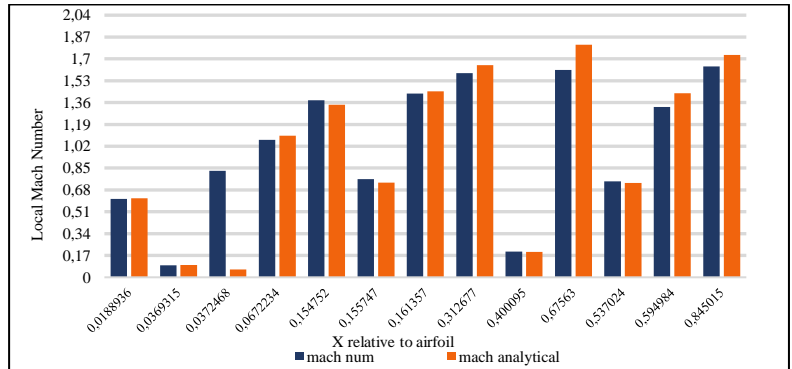

**Fig. 7.** Velocity evolution around the NACA 43013 wing airfoil, simulation for $M_\infty$=1.45

### 5.2   Statistical/Analytical Formulation of Empirical Parameter *m*

We have the incompressible Mach *m* that varies; according to x and y coordinates taken in different regions of the wall of the NACA 43013 wing profile as well as in different positions of $\bar{\eta}$ (characterizing the lower, intermediate and upper deck of the boundary layer), for a fixed upstream infinite Mach number $M_\infty$=*1.45*; according to a well-defined variable curve which is not linear (see figure 8).

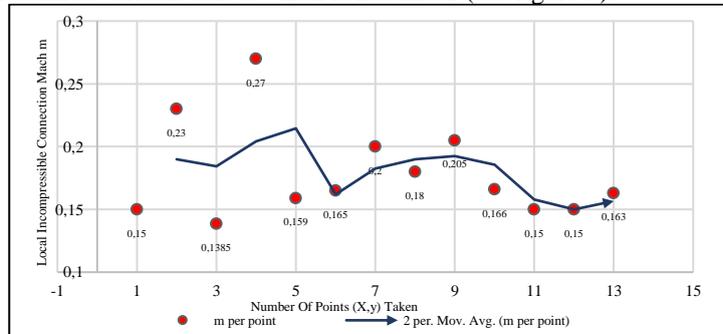

**Fig. 8.** Evolution of empirical parameter m for the points (x, y) taken

Thus, by plotting the incompressible Mach number *m* in a 3D Wireframe contour on EXCEL according to x and y coordinates for all chosen points, it's clear that *m* varies in quasi-linear way according to each parameter x and y individually (see figure 9) with *m* varying approximately between 0.1 and 0.2 (*maximum density in orange*) (which is consistent with our analytical results *0.152<m<0.2*).

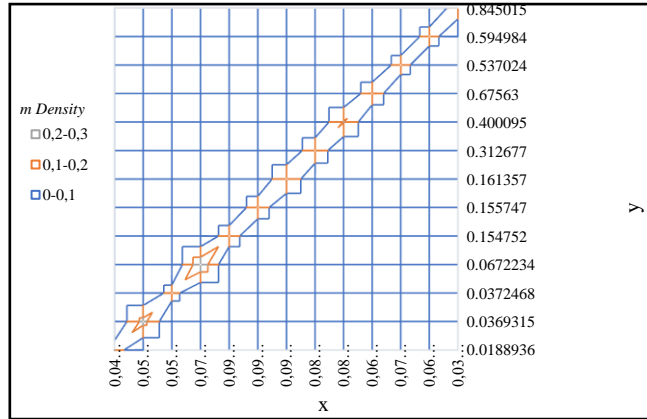

**Fig. 9. Wireframe Contour graph of the evolution of *m* density following (x, y) for all points taken for *M*∞=1.45 for NACA 43013**

And so, for a first determination of the value of m, we plot the data of the incompressible Mach *m* according to the coordinate points x and y at *M*∞=1.45 on the SPSS software, then we plot its boxplot, and we note that *m* is approximately equal to *m=0.164912* (the value of the median of the curve of *m* boxplot) (see figure 10).

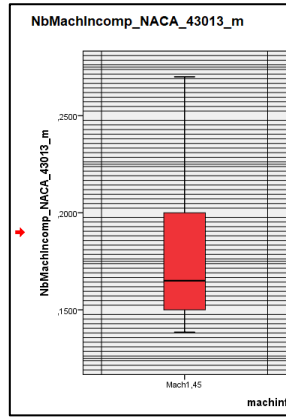

**Fig. 10. Boxplot graph on SPSS software for *m* as a function following points (x, y) for *M*∞=1.45 for NACA 43013**

Then for more precision for the *m*, we note thanks to the numerical simulation on Ansys Fluent that *m* depends mainly on three parameters for each specific airfoil $(\bar{y}, \bar{x}, M_\infty)$, from which, for more precision and to have exact results of our theoretical model, we assume that *m* varies for NACA 43013 according to these three parameters according to an empirical distribution function with:

$$m(\bar{y}, \bar{x}, M_\infty) = \alpha_1 \bar{y} + \alpha_2 \bar{x} + \alpha_3 M_\infty \quad \text{where } \alpha_1, \alpha_2 \ \& \ \alpha_3 \text{ are constants.} \quad (24)$$

For *M*∞=1.45 and NACA43013, we combine the calculated data into a single simple system of three equations:

$$\begin{cases} 0.0188936\alpha_2 + 0.0456681\alpha_1 + 1.45\alpha_3 = 0.15 \\ 0.0369315\alpha_2 + 0.0590449\alpha_1 + 1.45\alpha_3 = 0.23 \\ 0.0372468\alpha_2 + 0.0598307\alpha_1 + 1.45\alpha_3 = 0.1385 \end{cases} \quad (25)$$

In order to deduce, with more precision, the equation of *m* as a function of $(\bar{y}, \bar{x}, M_\infty)$, we solve the system (33) on the LINGO Software, we obtain the equation below which is the approximative statistical-analytical formulation of the local incompressible deck-connection Mach number *m* for $M_\infty$=1.45 for the NACA 43013 airfoil:

$$m_{NACA43013}(\bar{y}, \bar{x}, M_\infty = 1.45) = -168.301688\bar{y} + 129.2466\bar{x} + 5.39408395 \quad (26)$$

Same for $M_\infty$=1.55, following the same statistical/analytical method we obtain an approximative formulation for empirical parameter *m* for NACA43013 as following:

$$m_{NACA43013}(\bar{y}, \bar{x}, M_\infty = 1.5) = -130.33688\bar{y} + 39.387003\bar{x} + 6.486107077 \quad (27)$$

Consequently, all these comparisons on various parameters of the flow finally prove that the numerical simulation transcribes in a coherent and precise manner the flow around the profile in accordance with the theoretical approach. Indeed, the differences between the theoretical and numerical models are excessively small while being precise. Finally, the numerical approach leads to very fine concrete results. Thus, numerical simulation is a powerful, precise and reliable tool whose implementation is, moreover, much less cumbersome than the theoretical approach. However, a numerical simulation remains a series of theoretical calculations whose goal is to get as close as possible to reality, without ever reaching it. Therefore, it remains to be determined a method that will allow the phenomena to be truly observed in real or similar conditions. The experimental approach seems to meet these criteria.

## 6 Confrontation of Experimental-Analytical-Numerical Results & Discussion

The experimental aspect of our research was already carried out in the previous studies of our research team for NACA43013 airfoil with two pressure taps in the extrados at 31% and 69% of the chord [1]. To confront and validate our analytical model via experimentation we compared the different results for two upstream Mach numbers $M_{Upstream} = 1.55$ and $M_{Upstream} = 1.45$. For our analytical model, we took into consideration the statistical/analytical formulas of the empirical parameter *m* for each upstream Mach number, to enhance even more our model precision, results are shown in the table below:

**Table 1. Experimental-Numerical-Analytical Local Mach numbers results for NACA43013 at 31%c and 69%c for $M_\infty = 1.55$ and $M_\infty = 1.45$**

| Mach Upstream | X | Y | Exp Local Mach | Numerical Local Mach | Analytical Local Mach |
|---|---|---|---|---|---|
| 1.45 | 0.31 | 0,094 | 1.6 | 1.58245 | 1.59057 |
| 1.45 | 0.69 | 0,045 | 2.21 | 1.8625 | 2.0125 |
| 1.55 | 0.31 | 0,094 | 1.61 | 1.62156 | 1.60822 |
| 1.55 | 0.69 | 0,045 | 2.39 | 1.9524 | 2.1103 |

By comparing the experimental and analytical results for local Mach number near the wall far from the leading edge for a supersonic irrotational laminar steady flow of a viscous compressible fluid around thin airfoil NACA 43013, we can conclude to two main results:

- First, our improved statistical/analytical model is validated and proven to be accurate and exacte comparing to numerical simulation and experimentation,
- Second, the injection of the empirical parameter $m$ gave more exactitude to our analytical model comparing to reality {experimentation}, and decreased the relative error comparing to our previous model, as following:
    - For $M_\infty = 1.45$, for our previous model the error at 31%c was 1.25% and for our improved model it decreased to 0.5929%, same at 69%c the error decreased to 8.9367% comparing to 16.289% for the previous model, minimizing the relative error by more than half
    - For $M_\infty = 1.55$, for our previous model the error at 31%c was 1.242% and for our improved model it decreased to 0.1106%, same at 69%c the error decreased to 11.7029% comparing to 20.502% for the previous model, minimizing the relative error by more than half

The large difference between the experimental, numerical and analytical results in the second pressure tap which is located at 69% of the chord (near the trailing edge), can be explained by the fact that the flow in reality is turbulent, thus in the part of the airfoil close to the trailing edge is a part of instability, contrary to what we presumed in our assumptions that the flow is laminar and irrotational.

## 7　　Conclusion

The benefit of the SWBLI modelling is highly significant, since it doesn't only describe with high precision and takes into consideration a multitude of parameters that influence the flow behavior around a large variety of NACA airfoils with relative curvature $\lambda = 1$ in supersonic field with upstream Mach number $1.41 \leq M_\infty \leq 1.6864$, but it also is the fundamental base for any future passive flow control that could lead into the amelioration of the different factors governing the reliability and the airworthiness of the aircraft's wings such as the aerodynamic coefficients. The integration of the empirical parameter $m$ (Local Incompressible Connection Mach Number) added more value and precision to our analytical model, both comparing to numerical simulation where our model was validated, and experimentation results confrontation that proved the improvement of our model's precision by elaborating an exact formula for the empirical parameter based on statistical/analytical methods. As perspectives for our study, our research team aims to furthermore elaborate an exact fully analytical model for the empirical parameter $m$ taking into consideration the parameters and variables that influence its evolution $m(\overline{x}, \overline{y}, M_\infty, \beta)$, introducing the parameter $\beta$ that refers to the airfoil type.

## Acknowledgements

The authors would like to express profound gratitude and respects to the late Professor Mohammed Hasnaoui, who passed away due to Covid 19 during this study, in December 2020. His outstanding morality, astonishing expertise in the field of fluid mechanics and asymptotical modelling, as well as supervising role were very enriching; we, as a team, are honoured to have been his pupils.

## References

[1] (El-Aajine, O., Eddegdag, N., Naamane, A., Radouani, M., & El Fahime, B. 2022). Asymptotic Modeling of a Viscous Laminar Flow Around Thin Airfoils: Resolution and Experimental Treatment in Case of Supersonic Flow.

[2] (Dolling, D. S. 2001). Fifty years of shock-wave/boundary-layer interaction research: what next? *AIAA journal*, *39*(8), 1517-1531.

[3] (Ben Dor, G., Igra, O., & Elperin, T. 2001). Handbook of Shock Waves, Vol. 2 Shock Wave Interactions and Propagation.

[4] (Ligrani, P. M., McNabb, E. S., Collopy, H., Anderson, M., & Marko, S. M. 2020). Recent investigations of shock wave effects and interactions. *Advances in Aerodynamics*, *2*(1), 1-23.

[5] (Hasnaoui, M., Naamane, A., & Akhmari, H. 2019). Asymptotic Modeling the Aerodynamic Coefficients of the NACA Airfoil. Modelling. Measurement and Control B, 88(2-4), 58-66.

[6] (Naamane, A., & Hasnaoui, M. 2019). Supersonic Flow around a Dihedral Airfoil: Modeling and Experimentation Investigation. International Journal of Aerospace and Mechanical Engineering, 13(6), 413-417.

[7] (Zeytounian, R. K. 2002). Singular Coupling and the Triple-Deck Model. *Asymptotic Modelling of Fluid Flow Phenomena*, 471-525.

[8] (Neiland, V. Y. 1969). Theory of laminar boundary layer separation in supersonic flow. *Fluid Dynamics*, *4*(4), 33-35.

[9] (Stewartson, K., & Williams, P. G. 1969). Self-induced separation. *Proceedings of the Royal Society of London. A. Mathematical and Physical Sciences*, *312*(1509), 181-206.

[10] (Zeytounian, R. K. (Ed.). 1986). *Les Modèles Asymptotiques de la Mécanique des Fluides 1*. Berlin, Heidelberg: Springer Berlin Heidelberg, 158-171

[11] (Zeytounian, R. K. 1991). *Mécanique des fluides fondamentale* (Vol. 4). Springer Science & Business Media, 207-2026

[12] (Smith, F. T., Brighton, P. W. M., Jackson, P. S., & Hunt, J. C. R. 1981). On boundary-layer flow past two-dimensional obstacles. *Journal of Fluid Mechanics*, *113*, 123-152.

[13] (Matak, L., & Nikolić, K. K. 2022). CFD Analysis of F-16 Wing Airfoil Aerodynamics in Supersonic Flow. In *The Science and Development of Transport—ZIRP 2021* (pp. 197-210). Springer, Cham.

[14] (Billig, F. S. 1967). Shock-wave shapes around spherical-and cylindrical-nosed bodies. *Journal of Spacecraft and Rockets*, *4*(6), 822-823.